\documentclass[letterpaper, paper,11pt]{IAA-AAS}

\input{epsf}

\usepackage{bm}
\usepackage{amsmath}
\usepackage{subfig}
\usepackage{overcite}
\usepackage{footnpag}

\PaperNumber{-01-01}

\def\al{\alpha}

 \def\frac#1#2{{\textstyle{{#1}\over
{#2}}}} 
\def\lsim{\mathrel{\rlap{\lower4pt\hbox{\hskip1pt$\sim$}}
\raise1pt\hbox{$<$}}}
\def\gsim{\mathrel{\rlap{\lower4pt\hbox{\hskip1pt$\sim$}}
\raise1pt\hbox{$>$}}} \def\sqr#1#2{{\vcenter{\vbox{\hrule
height.#2pt \hbox{\vrule width.#2pt height#1pt \kern#1pt \vrule
width.#2pt} \hrule height.#2pt}}}}

\def\beq{\begin{equation}} \def\eeq{\end{equation}}
\def\beqa{\begin{eqnarray}} \def\eeqa{\end{eqnarray}}

\long\def\symbolfootnote[#1]#2{\begingroup
\def\thefootnote{\fnsymbol{footnote}}\footnote[#1]{#2}\endgroup}

\begin{document}

\title{THE CONTRIBUTION OF THERMAL EFFECTS TO THE ACCELERATION OF THE DEEP-SPACE PIONEER SPACECRAFT}

\author{Orfeu Bertolami\thanks{Departamento de F\'{\i}sica e Astronomia, Faculdade de Ci\^encias, Universidade do Porto, Rua do Campo Alegre 687, 4169-007, Porto, Portugal and Instituto de Plasmas e Fus\~ao Nuclear, Instituto Superior T\'ecnico, Universidade T\'ecnica de Lisboa, Av.\ Rovisco Pais 1, 1049-001 Lisboa, Portugal.}, Frederico Francisco\thanks{Instituto de Plasmas e Fus\~ao Nuclear, Instituto Superior T\'ecnico, Universidade T\'ecnica de Lisboa, Av.\ Rovisco Pais 1, 1049-001 Lisboa, Portugal.},
Paulo J. S. Gil\thanks{Departamento de Engenharia Mec\^anica and IDMEC - Instituto de Engenharia Mec\^anica, Instituto Superior T\'ecnico, Universidade T\'ecnica de Lisboa, Av.\ Rovisco Pais 1, 1049-001 Lisboa, Portugal.}
\ and Jorge P\'aramos\thanks{Instituto de Plasmas e Fus\~ao Nuclear, Instituto Superior T\'ecnico, Universidade T\'ecnica de Lisboa, Av.\ Rovisco Pais 1, 1049-001 Lisboa, Portugal.}
}

\maketitle


\begin{abstract}
	
A method for the computation of the radiative momentum transfer in the Pioneer 10 \& 11 spacecraft due to the diffusive and specular components of reflection is presented. The method provides a reliable estimate of the thermal contribution to the acceleration of these deep space probes and allows for a Monte-Carlo analysis from which an estimate of the impact of a possible variability of the parameters. It is shown that the whole anomalous acceleration can be explained by thermal effects. The model also allows one to estimate the expected time evolution of the acceleration due to thermal effects. The issue of thermal conduction between the different components of the spacecraft is discussed and confirmed to be negligible.

\end{abstract}




\section{Introduction}

The Pioneer anomaly (PA) has been an open issue in physics for over a decade: it consists of an presumably constant sun-bound acceleration on the Pioneer 10 and 11 deep-space probes, first put forward in a 1998 work by a team from the Jet Propulsion Laboratory (JPL) \cite{Anderson1998} and further scrutinized in a second paper, which settled  the anomaly as a cosntant acceleration $a_{\rm{Pio}} = (8.74 \pm 1.33) \times 10^{-10} ~ \rm{m/s^2}$ \cite{Anderson2002}. Independent data analyses \cite{Markwardt2002,Levy2009,Toth2009} have confirmed the existence of the anomalous acceleration, with at least two of them presenting results consistent with a non-constant acceleration \cite{Markwardt2002,Toth2009}.

The PA has attracted much attention from the scientific community throughout the last decade. Solutions range from conventional \cite{Katz1999, Scheffer2003} to new physics explanations \cite{Bertolami2004, Reynaud2005, Moffat2006, Bertolami2007}. For instance, it has been shown that the Kuiper Belt cannot be responsible for the anomalous acceleration \cite{Anderson2002,Bertolami2006}.

Rather surprisingly, the account of systematic effects presented in Ref.\ \cite{Anderson2002} dismissed altogether a significant contribution due to thermal effects. In opposition, it has been argued, albeit on a qualitative basis, that the on-board thermal power could indeed account for the anomalous acceleration \cite{Katz1999, Scheffer2003}.

Thus, the need for a quantitative description of the thermal effects became evident. Three independent efforts have been undertaken for the past few years, with the first results being released in 2008 by the Lisbon team. That work was based on the distribution of point-like Lambertian and isotropic radiation sources, validated by a set of test cases \cite{Bertolami2008,Bertolami2010}. The results suggest that thermal emissions of the spacecraft itself \cite{Bertolami2008,Bertolami2010}  could explain most if not all of the observed acceleration. These results have been confirmed by the finite-element modelling by the ZARM team \cite{Rievers2009,Rievers2010}. It has also been reported that an analysis is underway by the JPL based team \cite{Toth2008}.

Here, we consolidate our previous work by presenting a review of the results previously obtained through a direct modelling of reflection, complementing the estimates based on surface reflectivity \cite{Bertolami2008,Bertolami2010} which, when fully accounted, can explain the whole PA \cite{Francisco2012}. In addition, we discuss the effects of heat conduction between the main components and carry out a parametric analysis in order to establish reliable bounds for the results, as also discussed in Ref. \cite{Francisco2012}.


\section{Radiative Momentum Transfer}


\subsection{Point-like Source Method}\label{sec:radmomtransf}

In the present study, it is paramount that the approach chosen allows for the quick and reliable analysis of different scenarios and contributions. Motivated by the limitations inherent to the characterization of the anomalous acceleration itself, we adopted an approach that maintains a high degree of computational flexibility and speed, as outlined in Ref.\ \cite{Bertolami2008}. This uncertainty extends to the fact that both a constant acceleration and a linearly decaying one are consistent with the telemetry data \cite{Markwardt2002,Toth2009}, with the inclusion of a so-called ``jerk term'' (the derivative of the acceleration) found to be compatible with the expected temporal variation of a recoil force due to heat generated on board, with a half-life of $\sim 50~yr$. This makes the hypothesis of a thermal origin for the PA as the main culprit for the anomalous acceleration, as it would inevitably lead to a decay with at least the same rate as the power available onboard, which is derived from two $Pu$ radio-thermal generators (RTGs) with a half-life of $88~yr$. Possible causes for an enhanced decay include {\it e.g.} degradation of thermo-couples, stepwise shutdown of some systems and instruments, {\it etc.} \cite{Markwardt2002}.

With all this in mind, we designed a method that keeps the main physical features under control and available for scrutiny. The possibility that this simplicity and transparency is achieved at the expense of accuracy is offset by a series of test cases that are carried out to demonstrate the robustness of the results \cite{Bertolami2008, Bertolami2010}, validating the approach and showing that the level of accuracy is still much higher than that of the characterization of the acceleration itself. Furthermore, the method is further put to the test in a parametric analysis of the problem that weighs the relative importance of the different parameters involved.

The modelling is based on a distribution of point-like thermal radiation sources to account for the overall emission of the spacecraft. The formulation of emission and reflection is made in terms of the Poynting vector-field. We thus begin with the vector-field descriptions for the radiation emitting surfaces, modelled as Lambertian sources. The time-averaged Poynting vector field for a Lambertian source located at $\mathbf{x}_0$ is

\begin{equation}
	\label{lambertian}
	\mathbf{S}(\mathbf{x})={W \cos \theta \over \pi ||\mathbf{x}-\mathbf{x}_0||^2}{\mathbf{x}-\mathbf{x}_0 \over ||\mathbf{x}-\mathbf{x}_0||},
\end{equation}

\noindent where $W$ is the emissive power and $\theta$ is the angle with the surface normal. The value of $\cos \theta$ is the normalized inner product between the unitary emitting surface normal $\mathbf{n}$ and the emitted ray vector $(\mathbf{x}-\mathbf{x}_0)$.

The amount of energy illuminating a given surface $E_{\rm ilum}$ can be obtained by computing the Poynting-vector flux through the surface:

\begin{equation}
	E_{\rm ilum} = \int \mathbf{S} \cdot \mathbf{n}_{\rm ilum}~ dA,
\end{equation}

\noindent where $\mathbf{n}_{\rm ilum}$ is the normal vector of the illuminated surface.

The thermal radiation (infrared radiation) illuminating a surface will yield a force on that surface. This force per unit area is the \emph{radiation pressure} $p_{\rm rad}$, given by

\begin{equation}
p_{\rm rad}={\mathbf{S} \cdot \mathbf{n}_{\rm ilum} \over c},
\end{equation}

\noindent that is, the energy flux divided by the speed of light. This result should be multiplied by a factor $\alpha$ that varies between $\al =1$ for full absorption and $\al = 2$ for full reflection, which allows for an estimate of the reflection (as assessed in Refs.\ \cite{Bertolami2008,Bertolami2010}). However, a more rigorous treatment of reflection is presented in the next sections, following Ref.\ \cite{Francisco2012}.

Integrating the radiation pressure on a surface, we obtain the exerted force

\begin{equation}
\label{force_integration}
\mathbf{F} = \int {\mathbf{S} \cdot \mathbf{n}_{\rm ilum} \over c} {\mathbf{S} \over ||\mathbf{S}||} dA.
\end{equation}

\noindent The interpretation of this integration may not always be straightforward: to obtain the force exerted by the radiation on the emitting surface, the integral must be taken over a closed surface encompassing the latter. Analogously, the force exerted by the radiation on an illuminated surface requires an integration surface that encompasses it.

Also, considering a set of emitting and illuminated surfaces implies the proper account of the effect of the shadows cast by the various surfaces, which is then subtracted from the estimated force on the emitting surface. One may then straightforwardly read the thermally induced acceleration,

\begin{equation}
\mathbf{a}_{\rm th}={\sum_i \mathbf{F}_i \over m_{\rm pio}}.
\end{equation}


\subsection{Reflection Modelling -- Phong Shading}\label{sec:phong}

We suggest that in order to accurately model the reflections caused by the geometric configuration of the Pioneer 10 and 11 probes \cite{Francisco2012}, one can adopt a method developed in the 1970's by Bui Tuong Phong at the University of Utah and published in his Ph.D.\ thesis \cite{Phong} and known as \emph{Phong Shading}. This includes a set of techniques and algorithms commonly used to render the illumination of surfaces in 3D computer graphics. The method comprises two distinct parts:

\begin{itemize}
\item A reflection model including diffusive and specular reflection, known as \emph{Phong reflection model};
\item An interpolation method for curved surfaces modelled as polygons, known as \emph{Phong interpolation}.
\end{itemize}

The Phong reflection model is based on an empirical expression that yields the illumination of a given point in a surface $I_p$ as

\begin{equation}
I_p=k_a i_a + \sum_{m \in \text{lights}} \left[k_d (\mathbf{l}_m \cdot \mathbf{n})i_d + k_s (\mathbf{r}_m \cdot \mathbf{v})^{\alpha} i_s \right],
\end{equation}

\noindent where $k_a$, $k_d$ and $k_s$ are the ambient, diffusive and specular reflection constants, $i_a$, $i_d$ and $i_s$ are the respective light source intensities, $\mathbf{l}_m$ is the direction of the light source $m$, $\mathbf{n}$ is the surface normal, $\mathbf{r}_m$ is the direction of the reflected ray, $\mathbf{v}$ is the direction of the observer and $\alpha$ is a ``shininess'' constant (the larger it is, the more mirror-like the surface is).

This method provides a simple and straightforward method to model the various components of reflection, as well as a fairly accurate accounting of the thermal exchanges between the surfaces on the Pioneer spacecraft. There is no fundamental distinction between the treatment of infrared radiation, relevant for the Pioneer problem, and visible light, for which the method was originally developed, as long as one allows for a wavelength dependence of the above constants.

Given the presented thermal radiation in subsection \ref{sec:radmomtransf}, the Phong shading methodology can be easily adapted from a formulation based on \emph{intensities} (energy per surface unit per surface unit of the projected emitting surface) to one based on the energy-flux per surface unit (the Poynting vector). This is achieved through the following expressions for the diffusive component of the reflection

\begin{equation}
	\label{diffusive_reflection}
	\mathbf{S}_{\text{rd}}(\mathbf{x},\mathbf{x}')={k_d |\mathbf{S}(\mathbf{x}')\cdot \mathbf{n}| \over \pi ||\mathbf{x}-\mathbf{x}'||^2} (\mathbf{n} \cdot (\mathbf{x}-\mathbf{x}')) {\mathbf{x}-\mathbf{x}' \over ||\mathbf{x}-\mathbf{x}'||},
\end{equation}

\noindent and the specular component

\begin{equation}
	\label{specular_reflection}
	\mathbf{S}_{\text{rs}}(\mathbf{x},\mathbf{x}')={k_s |\mathbf{S}(\mathbf{x}')\cdot \mathbf{n}| \over {2 \pi \over 1+ \alpha} ||\mathbf{x}-\mathbf{x}'||^2} (\mathbf{r} \cdot (\mathbf{x}-\mathbf{x}'))^{\alpha} {\mathbf{x}-\mathbf{x}' \over ||\mathbf{x}-\mathbf{x}'||}.
\end{equation}

\noindent In both cases, the reflected radiation field depends on the incident radiation field $\mathbf{S}(\mathbf{x}')$ ($\mathbf{x}'$ is a point on the reflecting surface) and on the reflection coefficients $k_d$ and $k_s$, respectively. Using Eqs.~(\ref{diffusive_reflection}) and (\ref{specular_reflection}), we can compute the reflected radiation field by adding up these diffusive and specular components. From the emitted and reflected radiation, the irradiation of each surface is computed and, from it, a calculation of the force can be performed through Eq.~(\ref{force_integration}). This formulation allows for the determination of the force on the whole spacecraft, accounting for radiation that is reflected and absorbed by the various surfaces, as well as that which is propagated into space.

In the actual modelling of the spacecraft, once the radiation sources are in place, the first step is to compute the emitted radiation field and the respective force exerted on the emitting surfaces. This is followed by the determination of which surfaces are illuminated and the computation of the force exerted on those surfaces by the radiation. At this stage, we get a figure for the thermal force without reflections. The reflection radiation field is then computed for each surface and subject to the same steps as the initially emitted radiation field, leading to a determination of thermal force with one reflection. The only limitation to the iterative extension of this method to multiple reflection are the numerical integration algorithms and available computational power. If necessary, each step can be simplified through a discretization of the reflecting surface into point-like reflectors.


\section{Pioneer Thermal Model}


\subsection{Model Features}

We developed a geometric model of the Pioneer spacecraft that is compatible with the approach outlined in the previous sections ({\it cf.} Ref.\ \cite{Bertolami2008}). It makes a compromise between detail and simplicity, includes the most important features of the Pioneer spacecraft, namely:

\begin{itemize}
\item The parabolic high-gain antenna;
\item The main equipment compartment behind the antenna;
\item Two RTGs, cylindrical in shape, each connected to the main compartment through a truss.
\end{itemize}

\begin{figure}
	\centering
	\epsfxsize=15cm
	\epsffile{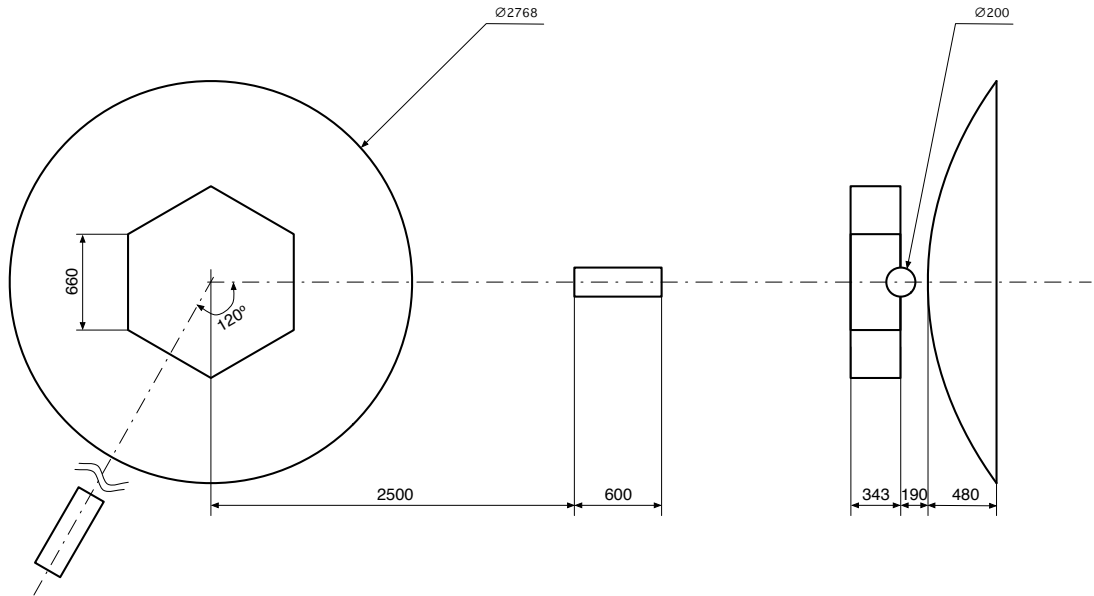}
	\caption{Schematics of the Pioneer geometric model used in our study, with relevant dimensions (in $\rm{mm}$); second RTG truss is not represented to scale. Lateral view indicates the relative position of the RTGs, box compartment and the gap between the latter and the high-gain antenna.}
	\label{pioneer_schematic}
\end{figure}

\noindent The geometric model with its respective dimensions is depicted in Fig.\ \ref{pioneer_schematic}. The model simplifies the minor surface features and details of the spacecraft. This simplification has been tested through specific test-cases discussed in Refs.\ \cite{Bertolami2008,Bertolami2010}, which show that its effect on the final result can be safely ignored for the purposes of this study.

The thermal radiation emissions modelling is achieved through a distribution of a few carefully placed point-like sources that mimic the actual emissions of the spacecraft as closely as possible. The fact that the Pioneer probes are spin-stabilised, considerably simplifies the problem, since the effect of all radial emissions is cancelled out in each full revolution of the spacecraft, leaving only contributions that are along the antenna's axis (here taken as the $z$-axis).

\begin{figure*}
	\subfloat[RTG]{
		\label{RTG_sources}
		\epsfxsize=7cm
		\epsffile{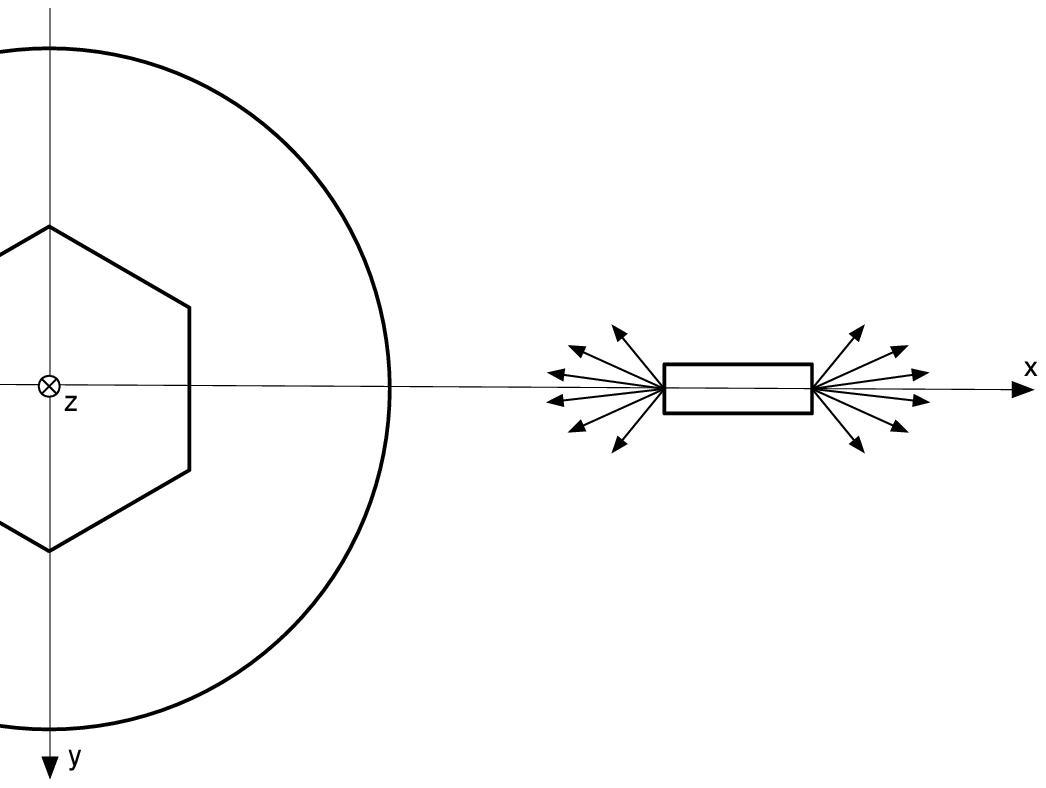}}
	\subfloat[Side walls of main compartment]{
		\label{equipment_sources}
		\epsfxsize=6cm
		\epsffile{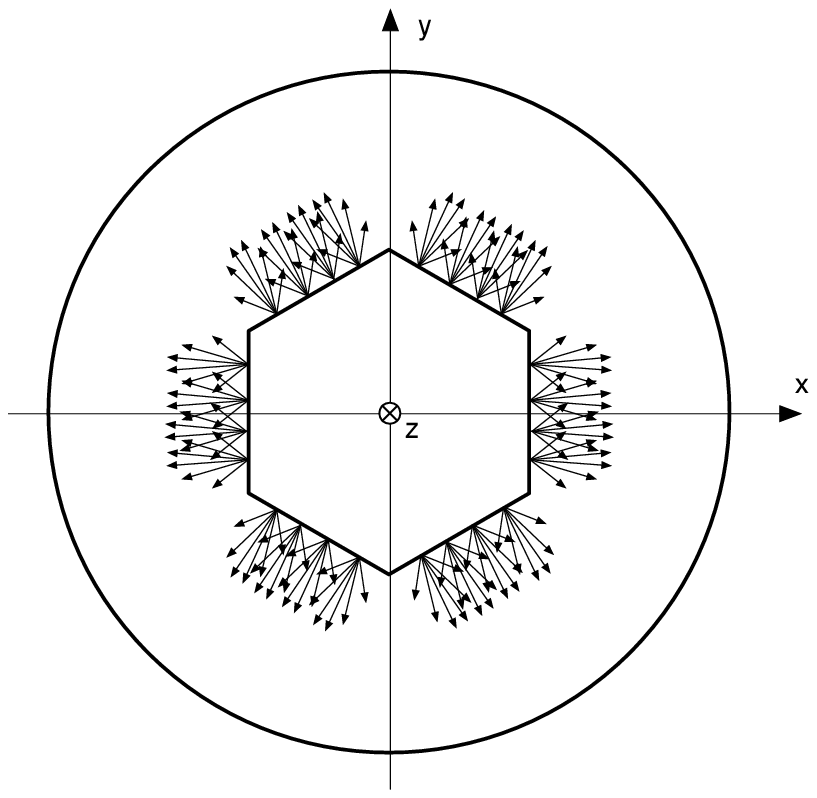}}
	\subfloat[Back wall of main compartment]{
		\epsfxsize=2.75cm
		\epsffile{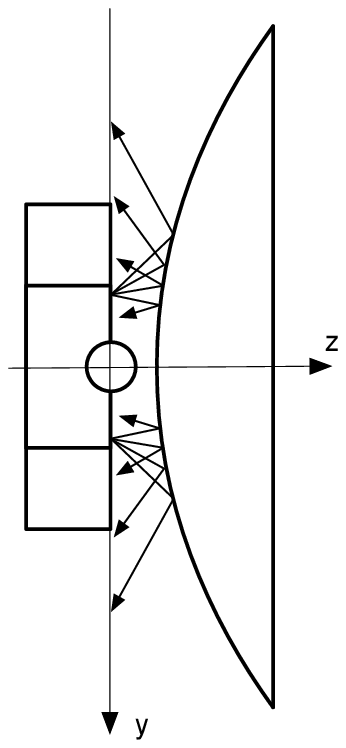}
		\label{back_wall}}
	\caption{Schematics of the Lambertian source distribution for the several radiative contributions.}
\end{figure*}

The main thermal radiation sources on the probe can be identified as the RTGs, where the main power source of the spacecraft is located, and the main equipment compartment, where the majority of the power is consumed. The RTGs can be easily and effectively modelled by two Lambertian sources, one at each base of the cylinder, as shown in Fig.~\ref{RTG_sources}. The emissions from the source facing outwards will radiate directly into space in a radial direction and its contribution will cancel-out. However, the radiation emitted towards the centre of the spacecraft will be reflected by both the high-gain antenna and the main equipment compartment.

The analysis of the main equipment compartment is divided between the front, back and lateral walls. The front wall of the latter (facing away from the sun and where the heat-dissipating louvers are located) will emit radiation directly into space, not illuminating any other surface. It can thus be modelled through a single radiation source, without impact on the final result.

The side walls of this compartment are each modelled by four Lambertian sources, as seen in Fig.~\ref{equipment_sources}. A previously conducted convergence analysis shows that this provides a reasonable degree of accuracy \cite{Bertolami2008,Bertolami2010}. This radiation will reflect mainly on the high-gain antenna.

Finally, we included the contribution of the back wall of the main equipment compartment (facing the high-gain antenna). The radiation from this wall will, in a first iteration, reflect off the antenna and add a contribution to the force in the direction of the sun, as depicted in Fig.~\ref{back_wall}. This back wall was modelled using a set of six Lambertian sources evenly distributed in the hexagonal shape. The relevant contributions for this analysis can be summarized in Table~\ref{pioneer_components}, with each of them indexed for reference in the following sections.

\begin{table*}
	\begin{center}
	\caption{Indexing of the components considered in this study}
	\begin{tabular}{c c c}
		\hline
		Emitting surface					& Reflecting surface					& Index \\
		\hline
		\hline
		Lateral surface of main compartment	& High-gain antenna dish				& 1.1 \\
		RTG									& High-gain antenna dish				& 2.1 \\
		RTG									& Lateral surface of main compartment	& 2.2 \\
		Back surface of main compartment	& High-gain antenna dish				& 3.1 \\
		Front surface of main compartment	& none									& 4 \\
		\hline
		\label{pioneer_components}
	\end{tabular}
	\end{center}
\end{table*}

A brief remark about the experimental setup section attached to one of the sides of the main compartment. The full set of scientific instruments uses 24.5 W, out of a total of 120 W of dissipated electrical power (after launch) \cite{Turyshev2008}. Although not negligible, the preceding discussion indicates that disregarding this difference in geometry should not change the outcome significantly. Naturally, a more detailed geometrical modelling of the spacecraft would lead to a more refined value for the overall thermal output, but this is beyond the scope of the current work. A more detailed discussion on the hypotheses used in the Pioneer model can be found in Refs.~\cite{Bertolami2008,Bertolami2010}.


\subsection{Conductive Contribution}

Since the components of the spacecraft are mechanically connected, wee should expect a certain amount of conductive heat flux between them, changing the amount of energy that each one radiates to maintain the equilibrium. The structural elements connecting different components of the spacecraft, namely, between the RTGs and the main compartment and the main compartment and the high gain antenna should be looked upon more closely. In both cases, considering the approximate thickness, conductivity of the truss bars and temperature gradients suggest that the effect of conduction between these components is always a small fraction (typically, a few Watts) of their respective powers.

Let us consider what we can expect from the conductive effect. A difference in temperature is expected between different spacecraft elements e.g.\ the RTGs and the main compartment. Neglecting conduction, they can be considered approximately in equilibrium, with the energy dissipated from the RTGs and electrical equipment radiated away from themselves. However, there will be a certain amount of conductive heat flow between them through the truss. This would correspond to a transfer of power from the hotter to the colder elements (from the RTG to the main compartment). Realistically, the transferred heat will tend to increase the temperature of the colder elements and decrease the temperature of the hotter ones, decreasing the conductive heat flux, until a new equilibrium. This effect can be taken into account by a variation of the relative power distribution between elements, and since it is rather small (see below), it can be dealt with in the Monte-Carlo calculations.

Focusing on the connection between the RTGs and the main compartment, where the effect of thermal conduction is expected to be greater, we can estimate the effect for a given reasonable difference of temperatures. Assuming a temperature near $0^\circ \rm C$ in the equipment compartment (a worst case scenario, since it is actually warmed by the electronics to $\sim 10^\circ \rm C$ \cite{temperature1,temperature2}) and the RTGs near $150^\circ \rm C$, a temperature gradient of approximately $60 ~\rm K/m$ is obtained. Each truss is composed of three small diameter rods, that we estimate to have a total cross-section of the order of $10^{-4}~\rm m^2$, made out of aluminium that has a conductivity of approximately $240~\rm W/(m \cdot K)$. These figures would translate into a total conducted power of the order of $1~W$ (up to $4~W$ in more conservative estimates), which is clearly negligible, since it is two orders of magnitude below the power of the main compartment and three orders below the RTG power.

The power conduction from the main compartment to the antenna would be even smaller, since the temperature gradient is also much smaller. This effect can then be safely disregarded in any further computations of the global effect and easily taken into account in the relative power distribution between spacecraft elements.


\subsection{Radiative and Reflective Contributions}

We can now compute the contribution of the individual components listed in Table \ref{pioneer_components}. This is done through the integration of Eq.~(\ref{force_integration}) in three successive steps. First, the emitted radiation field given by Eq.~(\ref{lambertian}) is integrated along a closed surface, yielding the first-order effect of the emissions. Afterwards, the same radiation field is integrated along the illuminated surfaces, in order to subtract the shadow effect. Finally, the reflected radiation vector-field, given by Eqs.~(\ref{diffusive_reflection}) and (\ref{specular_reflection}), is integrated along closed surfaces, adding the contribution from reflection. This process allows us to obtain the values for the force in terms of the emitted powers and reflection coefficients. As already pointed out, the results that follow are along the main antenna axis, since all radial components cancel-out. A positive figure indicates a sunward force.

The computation of the contribution from the front surface of the main compartment (index 4) is fairly straightforward, since there are no reflections on other surfaces. For this reason, and the fact that this surface is perpendicular to the spacecraft's spin axis, it is effectively modelled by a single radiation source, as indicated in Table~\ref{sources}. by replacing the position and surface normal direction in Eq.~(\ref{lambertian}) we obtain the emitted radiation field. The force exerted by the radiation field on the emitting  surface is obtained by integrating Eq.~(\ref{force_integration}) along a closed surface --- in this case, chosen as a half-sphere centered at the location of the radiation source. The $z$ component of the resulting force on the emitted radiation is, as expected, given by

\begin{equation}
F_4 = {2 \over 3} {W_{\rm front} \over c}.
\end{equation}

\begin{table*}
	\begin{center}
	\caption{Position and direction of the Lambertian source used to model each emitting surface of the Pioneer spacecraft model.}
	\begin{tabular}{c c c c}
		\hline
		Emitting Surface	& Source	& Position (m)	& Surface Normal (m) \\
		\hline
		\hline
		Front wall (index 4)& 1 		& $(0,0,-0.343)$	& $(0,0,-1)$ \\
		\hline
		Lateral wall		& 1 		& $(0.572,0.2475,-0.172)$	& $(1,0,0)$ \\
		(index 1.1)			& 2 		& $(0.572,0.0825,-0.172)$	& $(1,0,0)$ \\
							& 3 		& $(0.572,-0.0825,-0.172)$	& $(1,0,0)$ \\
							& 4 		& $(0.572,-0.2475,-0.172)$	& $(1,0,0)$ \\
		\hline
		RTG					& 1 		& $(2.5,0,0)$	& $(-1,0,0)$ \\
		(index 2.1 \& 2.2)	& 2 		& $(3.1,0,0)$	& $(1,0,0)$ \\
		\hline
		Back wall			& 1 		& $(0.381,0,0)$	& $(0,0,1)$ \\
		(index 3.1)			& 2 		& $(0.191,0.33,0)$	& $(0,0,1)$ \\
							& 3 		& $(-0.193,0,33)$	& $(0,0,1)$ \\
							& 4 		& $(-0.381,0,0)$	& $(0,0,1)$ \\
							& 5 		& $(-0.191,-0.33,0)$	& $(0,0,1)$ \\
							& 6 		& $(0.191,-0.33,0)$	& $(0,0,1)$ \\
		\hline
		\label{sources}
	\end{tabular}
	\end{center}
\end{table*}

The radiation emitted from the lateral walls of the main compartment do mainly illuminate the high-gain antenna (index 1.1). Due to the already discussed axial symmetry of the problem, and neglecting the interaction with the small and far away RTGs, it is only necessary to model one of the six walls. The set of Lambertian sources used for one of these walls is indicated in Table~\ref{sources}. The $z$ component of the radiation field force on the emitting surface vanishes, as the emitting surface is orthogonal to the $z$-axis. Integrating Eq.~(\ref{force_integration}) over the illuminated portion of the antenna dish, we get the force exerted on the illuminated surface, which accounts for the shadow effect. This gives a $z$ component of $-0.0738(W_{\rm lat} / c)$, where $W_{\rm lat}$ is the power emitted from the lateral walls --- to be subtracted from the total force of the emitted radiation.

The computation of diffusive reflection is made through Eq.~(\ref{diffusive_reflection}), that returns the reflected Poynting vector-field $\mathbf{S}_{\text{rd}}(\mathbf{x},\mathbf{x}')$ due to the emitted radiation field $\mathbf{S}(\mathbf{x}')$, where $\mathbf{x}'$ is a point in the reflecting surface. The reflected radiation field is given for each point in the reflecting surface and must be integrated first over the reflecting surface itself, conveniently parameterized. This gives the overall reflected radiation field that is then integrated through Eq.~(\ref{force_integration}) over a closed surface to obtain the force resulting from the reflected radiation. The procedure for specular reflection is entirely similar, except that Eq.~(\ref{specular_reflection}) should be used to obtain the reflected radiation field before the integration.

Integrating the vector-field representing radiation from the lateral walls of the main compartment reflecting on the high-gain antenna, we obtain a force result of $0.0537 k_{\rm d,ant}(W_{\rm lat} / c)$ for the diffusive component and $0.0089 k_{\rm s,ant}(W_{\rm lat} / c)$ for the specular component, where $W_{\rm lat}$ is the power emitted from the referred walls and $k_{\rm d,ant}$ and $k_{\rm s,ant}$ are the diffusive and specular reflection coefficients of the main antenna, respectively.

The result for the contribution is given by adding the (vanishing) emitted radiation force, the shadow effect and both components of reflection, leading to

\begin{equation}
F_{11} = {W_{\rm lat} \over c} (0.0738 + 0.0537 k_{\rm d,ant} + 0.0089 k_{\rm s,ant}).
\end{equation}

The emissions from the RTGs were modelled by a Lambertian source at each base of each cylindrical shape RTG, as listed in Table~\ref{sources}. Similarly to the lateral walls, only the modelling of one RTG is required, since the effect of the radial components cancel-out with each revolution of the probe. It can also be easily shown that only the emissions from the base facing the centre of the spacecraft (source 1 of the RTG in Table~\ref{sources}) will impact on the acceleration along the $z$-axis. Emissions from the base facing outwards (source 2) are not reflected on any surface and its contribution vanishes when averaged over each revolution of the spacecraft.

Using the same procedure, the generated force from the RTG emissions is thus given in terms of the power emitted from the RTG bases facing the centre of the spacecraft $W_{\rm RTGb}$. The force resulting from reflections on the antenna (index 2.1) is given by

\begin{equation}
F_{21} = {W_{\rm RTGb} \over c} (0.0283 + 0.0478 k_{\rm d,ant} + 0.0502 k_{\rm s,ant}),
\end{equation}

\noindent and the contribution from reflections on the lateral surfaces of the main equipment compartment is

\begin{equation}
F_{22} = {W_{\rm RTGb} \over c} (-0.0016 + 0.0013 k_{\rm s,lat}),
\end{equation}

\noindent where $k_{\rm d,ant}$, $k_{\rm s,ant}$, $k_{\rm d,lat}$ and $k_{\rm s,lat}$ are the respective reflection coefficients.

As previously argued \cite{Bertolami2008}, the contribution from radiation emitted from the back wall of the main compartment and reflecting in the space between this compartment and the antenna dish would be small. In an attempt to confirm this assumption, a computation was made using the method described above. The results ultimately show that this contribution cannot be overlooked. Considering one reflection from the antenna dish, the result in terms of the emitted power from the back wall of the main compartment $W_{\rm back}$, by

\begin{equation}
F_{3} = {W_{\rm back} \over c} \left( -{2 \over 3} + 0.5872 + 0.5040 k_{\rm d,ant} + 0.3479 k_{\rm s,ant} \right).
\end{equation}
Note that the $-{2 \over 3}{W_{\rm back} \over c}$ is the contribution from the emitted radiation and $0.5872 {W_{\rm back} \over c}$ is the effect of the antenna's shadow.

From the force computations, once the respective powers and reflection coefficients are inserted, the final result of the acceleration due to thermal dissipation mechanism follows:

\begin{equation}
a_{\rm th} = {F_{11} + F_{21} + F_{22} + F_{3} +F_{4} \over m_{\rm Pio}},
\end{equation}

\noindent where the mass of the spacecraft is taken at an approximate value $m_{\rm Pio} = 230~\rm{kg}$. This figure considers a total mass of $259~\rm{kg}$ at launch, including $36~\rm{kg}$ of hydrazine propellant that was partially consumed in the early stages of the mission \cite{Anderson2002}. Note that this is an approximate figure, since the actual masses for the Pioneer 10 and 11 would be slightly different due to different fuel consumptions.


\subsection{Available Power}
\label{available_power}

It was chosen from the earliest stages of this study to use the available onboard power as the independent variable in the computation of the thermally induced acceleration. This is justified due to the reasonably good knowledge of the available power --- indeed, this is one of the few parameters with reliable data available throughout the operational lifetime of the probes --- and also because power is the driving parameter determining the emitted thermal radiation.

The power on board the Pioneer probes comes from the two plutonium RTGs. It is thus easy to compute the total power available, considering the $87.74~\rm{year}$ half-life of plutonium. According to Ref. \cite{Anderson2002}, the total thermal power of the RTGs at launch was $2580~\rm{W}$, leading to a time evolution given by

\begin{equation}
	W_{\rm tot} = 2580 \exp \left(- {t \ln 2 \over 87.72} \right)~\rm{W},
	\label{total_power}
\end{equation}

\noindent with $t$ being the time in years from launch.

Electrical power generation is ensured by a set of thermocouples located in the RTGs. Most of this power is consumed by the various systems and instruments located in the main equipment compartment, except for a small portion used by the radio signal. A good measurement of the electrical power is available through telemetry data \cite{Toth2008}. Knowing the total electrical power consumption, remaining unused power is mostly dissipated at the RTGs themselves, through its external surface and radiating fins.

At launch, $120~\rm{W}$ of electrical power were being delivered to the main equipment compartment plus around $20~\rm{W}$ for the radio transmission to Earth, leaving $2420~\rm{W}$ of thermal power in the RTGs. It is also known from telemetry data that the electrical power decayed at a faster rate than thermal power, with its half-life being around $24~\rm{years}$. This would lead to an approximate time evolution of the electrical power in the equipment compartment given by

\begin{equation}
	W_{\rm equip} \approx 120 \exp \left(- {t \ln 2 \over 24} \right)~\rm{W},
	\label{elec_power}
\end{equation}

\noindent which is consistent with Fig.~11 in Ref.~\cite{Toth2008}.

The above considerations and the power values extracted from the available telemetry data for the latest stages of the mission --- specifically, the reading for the twenty six years after launch (for the Pioneer 10, up to 1998) --- are used as a baseline scenario. In a second stage of this study, the time evolution is taken into account, according to the reasoning developed in this section.


\section{Results and Discussion}


\subsection{Parametric Analysis and the Effect of Conduction}

We now perform a static parametric analysis, in an attempt to establish a reliable estimate for the thermal acceleration at an instant 26 years after launch. The analysis resorts to a classic Monte-Carlo method, where a probability distribution is assigned to each variable and random values are then generated. A distribution of the final result ({\it i.e.} the acceleration) is then obtained.

We establish a set of scenarios for the distribution of the emitted power throughout the different surfaces, while keeping the total power constant as $W_{\rm tot} = 2100~\rm{W}$ and the electrical power as $W_{\rm equip} = 56~\rm{W}$, leaving RTG thermal power at $W_{\rm RTG} = 2024~\rm{W}$ (assuming the power of the radio beam is still $20~\rm{W}$). These scenarios act as a baseline for the parametric analysis and are summarised in Table \ref{baseline_results}.

The parameters that come into play in this setup are the power emitted from each surface, $W_{\rm RTGb}$, $W_{\rm front}$, $W_{\rm lat}$, $W_{\rm back}$, and the reflection coefficients $k_{\rm d,ant}$, $k_{\rm s,ant}$ and $k_{\rm s,lat}$. A quick analysis of Table \ref{baseline_results} allows us to draw a qualitative assessment: for instance, the of power emitted from the front wall $W_{\rm front}$ has a decisive influence in the final result. In contrast, the relevance of the specular reflection coefficient of the lateral wall $k_{\rm s,lat}$ is almost negligible.

\begin{table*}[ht]
	\begin{center}
	\caption{Pioneer thermal acceleration results for baseline scenarios (1: Lower bound, uniform temperature; 2: Higher emissions from louvers; 3: Diffusive reflection in antenna; 4: Diffusive and specular reflection; 5: Upper bound).}
	\begin{tabular}{c | c c c c | c c c | c}

 Scenario	& $W_{\rm RTGb}$	& $W_{\rm front}$	& $W_{\rm lat}$	& $W_{\rm back}$	& $k_{\rm d,ant}$	& $k_{\rm s,ant}$	& $k_{\rm s,lat}$	& $a_{\rm th}$ \\
		 & $(\rm{W})$		& $(\rm{W})$		& $(\rm{W})$	& $(\rm{W})$		&				&				&				& $(10^{-10}~\rm{m/s^2})$ \\
		\hline
		\hline
		1	&$143.86$	&$17.5$	&$21$	&$17.5$	&$0$ 	&$0$ 	&$0$	& $2.27$ \\
		2	&$143.86$	&$40$	&$8.73$	&$7.27$	&$0$ 	&$0$ 	&$0$	& $4.43$ \\
		3	&$143.86$	&$40$	&$8.73$	&$7.27$	&$0.8$ 	&$0$ 	&$0$	& $5.71$ \\
		4	&$143.86$	&$40$	&$8.73$	&$7.27$	&$0.6$ 	&$0.2$ 	&$0.4$	& $5.69$ \\
		5	&$158.24$	&$56$	&$0$	&$0$	&$0.8$ 	&$0$ 	&$0.4$	& $6.71$ \\
		
	\end{tabular}
	\label{baseline_results}
	\end{center}
\end{table*}

For the static analysis at $t=26~\text{years}$, Scenario 4 is taken as a baseline, since it is the one more solidly based on physical arguments. The power emitted by the RTG bases facing the main compartment $W_{\rm RTGb}$ is generated from a Gaussian distribution with the mean value of $143.86~\rm{W}$ and a standard deviation of $25\%$ of this value. This allows for a significantly larger deviation than that considered in the top-bound scenario (Scenario 5), which had only a $10\%$ increase in the power of this surface. This is made so to account for any unanticipated anisotropies in the temperature distribution of the RTGs and enhance the confidence of the results.

In what regards the main equipment compartment, the focus is on the power emitted by the louvers located in the front wall. The selected distribution for the parameter $W_{\rm front}$ is also normal, with the mean value at $40~\rm{W}$ (also corresponding to Scenario 4 and the standard deviation at $7.5~\rm{W}$, so that the 95\% probability interval ($2 \sigma$) for the value of $W_{\rm front}$ is below the top figure of $56~\rm{W}$, which corresponds to all the power from the equipment being dissipated from the front wall. For the remaining surfaces of the equipment compartment, the power is computed at each instance so that the total power of the equipment is maintained at $56~\rm{W}$.

The reflection coefficients for the antenna are allowed to take values in an uniform distributions bounded at $[0.6,0.8]$ for $k_{\rm d,ant}$ and $[0,0.2]$ for $k_{\rm s,ant}$, with the condition $k_{\rm d,ant}+k_{\rm s,ant}=0.8$, since this is the reference value for aluminium in infrared wavelengths. We also expect the specular component to be small, since the surface is not polished. Furthermore, if we allow for the possibility of surface degradation with time along the mission, the specular component would suffer a progressive reduction in favour of the diffusive component, a possibility that this analysis takes into account.

We performed $10^4$ Monte Carlo iterations, which easily ensures the convergence of the result. The thermal acceleration estimate yielded by the simulation for an instant $26~\text{years}$ after launch, with a $95\%$ probability, is

\begin{equation}
	a_{\rm{th}}(t=26)=(5.8 \pm 1.3) \times 10^{-10} ~ \rm{m/s^2}.
\end{equation}

\begin{figure}
	\centering
	\epsfxsize=10cm
	\epsffile{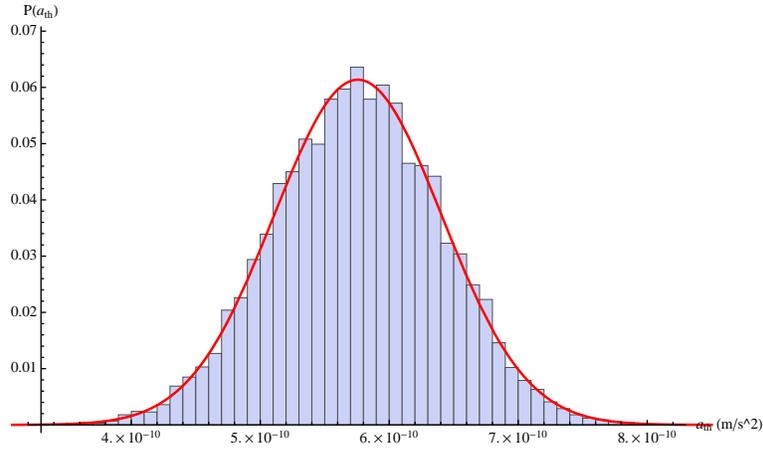}
	\caption{Histogram for the probability distribution resulting from the Monte Carlo simulation with 10000 iterations for the thermal acceleration of the spacecraft at $t=26~\text{years}$ after launch, with the normalised Gaussian distribution superimposed.}
	\label{static_hist}
\end{figure}

\noindent This result is extracted from the approximately normal distribution shown in Fig.~\ref{static_hist}. The conformity of the results to a normal distribution was confirmed by a Shapiro-Wilk normalcy test with a $p$-value $\sim 0$.

These results represent $44\%$ and $96\%$ of the reported value $a_{\rm{Pio}}=(8.74 \pm 1.33) \times 10^{-10}~\rm{m/s^2}$ which was obtained under the assumption that it was constant acceleration. At the very least, this strongly suggests a preponderant contribution of thermal effects to the PA.

We can now repeat the parametric analysis, but assuming that heat conduction takes place. In this case, we adjust the power distribution so that around $4~\rm W$ of thermal power are being transferred from the RTGs to the main compartment and run the simulation with the same 10000 iterations.

\begin{figure}
	\centering
	\epsfxsize=10cm
	\epsffile{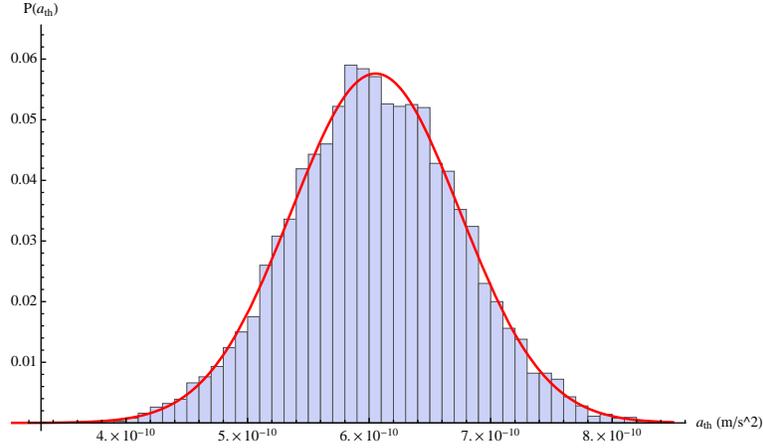}
	\caption{Histogram for the probability distribution resulting from the Monte Carlo simulation with 10000 iterations for the thermal acceleration of the spacecraft at $t=26~\text{years}$ after launch considering heat conduction between the RTGs and main compartment, with the normalized Gaussian distribution superimposed.}
	\label{static_hist_cond}
\end{figure}

The results show a slight increase in the overall acceleration of around $5\%$, {\it i.e.},
\begin{equation}
	a_{\rm{th}}(t=26,{\rm cond})=(6.05 \pm 1.4) \times 10^{-10} ~ \rm{m/s^2}.
\end{equation}
The resulting distribution is shown in Fig.~\ref{static_hist_cond}. This estimate represents an upper bound for the conductive heat transfer, confirming that this effect does not have a dramatic impact on the final result and does not alter any conclusion that one may or may not draw from them.

\subsection{Time Evolution}

The last issue of our analysis concerns the expected time evolution of the thermal acceleration experienced by the Pioneer 10 and 11 spacecraft.

\begin{figure}
	\centering
	\epsfxsize=12cm
	\epsffile{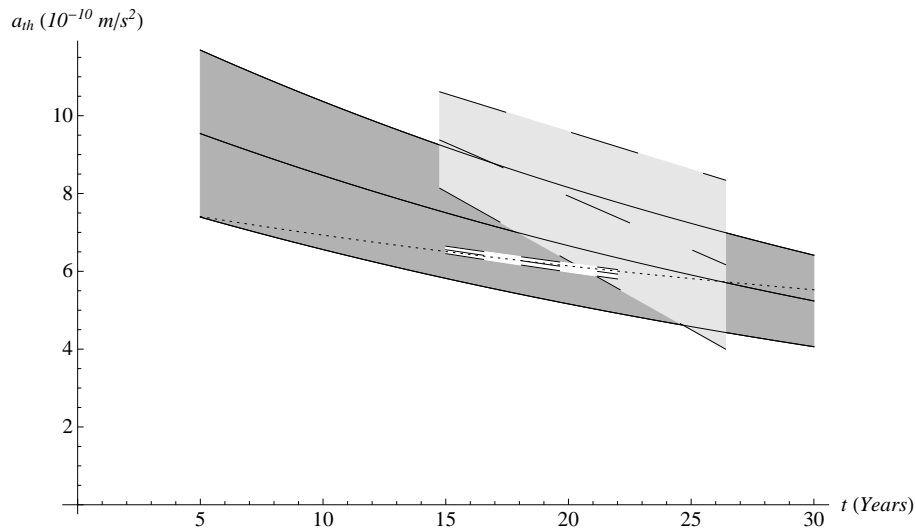}
	\caption{Results for the time evolution of the thermal acceleration on the Pioneer spacecraft compared with results based on two data analyses with non-constant solutions for the anomalous acceleration. The dotted line is the time extrapolation of the static analysis of the thermal acceleration and the dark grey area correspond to a $95\%$ probability for the thermal acceleration in the time evolution analysis. For comparison, the light grey area is based on results from the data analysis of Refs.~\cite{Toth2009} and \cite{Markwardt2002}, respectively.}
	\label{time_evolution}
\end{figure}

A first estimate can be obtained by extrapolating the static results with the time evolution of electric power, using Eqs. (\ref{total_power}) and (\ref{elec_power}). Results of this extrapolation are shown as the dotted line in Fig.~\ref{time_evolution}, with the approximate exponential decay of the available power translated into a similar trend in the evolution of the thermally induced acceleration. This extrapolation, however, does not account for the possible temporal variation of some parameters --- particularly, the power distribution throughout the different surfaces or their reflection coefficients. This could be accounted by a simulation of the full span of the missions ({\it i.e.} a large number of consecutive simulations), with a specific prescription for the variability of these parameters.

Such task will be addressed in the future. For now, we have chosen a somewhat simpler approach to grasp of the possibility discussed above: we apply the Monte-Carlo static analysis to only two earlier moments of the mission. Each simulation produces a central value, with top and lower bounds that are then fitted to an exponential trend, leading to an estimated time evolution of the thermal acceleration. The chosen instants for the earlier static analysis were at $t=8 ~\text{years}$ and $t=17 ~\text{years}$ after launch, corresponding, respectively, to the 1980 and 1989 values for the Pioneer 10. The 1980 date is approximately when the effect of the solar radiation pressure dropped below $5 \times 10^{-10}~\rm{m/s^2}$ \cite{Anderson2002}.

The procedure is similar to the analysis of the previous subsection, but using the 1980 and 1989 available power values as a base for the choice of the distributions. The resulting thermal acceleration is, for $t=8 ~\text{years}$,

\begin{equation}
a_{\rm{th}}(t=8)=(8.9 \pm 2) \times 10^{-10} ~ \rm{m/s^2},
\end{equation}

\noindent corresponding to the same $95\%$ probability, and for $t=17 ~\text{years}$

\begin{equation}
a_{\rm{th}}(t=17)=(7.1 \pm 1.6) \times 10^{-10} ~ \rm{m/s^2}.
\end{equation}

Using the three static estimates presented above, it is now possible to produce a time evolution based on a fit to an exponential decay. This is performed for the mean value, top-bound and lower-bound of the acceleration, always based on a $95\%$ probability degree. The curve fit for the mean, upper and lower values of the thermal acceleration reads

\begin{equation}
a_{\rm th} =  [(1.07 \pm 0.24) \times 10^{-9}] \exp (-0.0240t)~\rm{m/s^2},
\end{equation}

\noindent with $t$ corresponding to the time after launch in $\text{years}$.

The time evolution resulting from any of these scenarios corresponds to a decay with a half-life of approximately $60~\text{years}$, related to the nuclear decay of the plutonium in the RTGs and the faster decay rate of the electrical power, already discussed in Section~\ref{available_power}. The graphic representation of the band of values predicted by our model is shown in Fig~\ref{time_evolution} (dark grey region) and compared with the values indicated by non-constant results for the anomalous acceleration in Refs.~\cite{Toth2009,Markwardt2002} (light grey region).

\section{Conclusion and Outlook}

The method developed to account for reflection on the surfaces of the Pioneer spacecraft allows for the modelling of thermal radiation with increased accuracy, while maintaining the desired simplicity and computational speed of our previously proposed  approach \cite{Bertolami2008}. This new tool allows for a successful modelling of the most relevant features of the Pioneer spacecraft concerning thermal effects and their impact on the resulting acceleration.

The results provided by the developed method, based on Phong shading, globally confirm those previously obtained in Refs.~\cite{Bertolami2008,Bertolami2010}~~: the acceleration resulting from thermal effects has an order of magnitude that is compatible with the constant anomalous acceleration reported in Ref.~\cite{Anderson2002}. We believe that this problem is especially suited to the chosen approach, considering its specific features. Furthermore, this Phong shading method is capable of being adapted for future studies of radiation momentum transfer in other spacecraft.

The main difficulty that this problem sets has always been the lack of sufficient and reliable data for a detailed engineering modelling of the spacecraft's flight conditions, despite various claims otherwise. We have overcome this barrier through a parametric analysis that takes into account a wide range of different scenarios. Indeed, this contribution clearly shows how an extra effect not previously taken into account, the thermal conduction between spacecraft elements, can be easily integrated in our model to account for its impact through the Monte-Carlo simulations. It turns out that this effect does not have a significant impact and does not change the overall conclusion. This further demonstrates the value of the chosen strategy that allowed us to present a range of probable values for the thermal effects, which apparently have a signature that is compatible with the Pioneer anomalous acceleration. This conclusion has also recently been confirmed by the team working in Bremen using a finite-element method \cite{Rievers2011}. This, we believe, brings us significantly closer to a solution for this problem in terms of onboard thermal effects.


\section{Aknowledgments}

The work of OB and JP is partially supported by the Funda\c{c}\~{a}o para a Ci\^{e}ncia e Tecnologia (FCT), through the project PTDC/FIS/111362/2009. The work of FF is sponsored by the FCT, under the grant BD 66189/2009.


\bibliographystyle{AAS_publication}   
\bibliography{Pio_Thermal_DyCoSS}   

\end{document}